# Rotating thermoelectric device in periodic steady state

I.V. Bezsudnov[1,2], A.A. Snarskii[3,4]

[1]Nauka – Service JSC, Moscow, Russia

[2]ITMO University, St. Petersburg, Russia

[3]Dep. of Gen. and Theor. Physics, National Technical University of Ukraine "KPI", Kiev, Ukraine

[4]Institute for Information Recording NAS Ukraine, Kiev, Ukraine

Abstract

We propose the rotating thermoelectric (TE) device comprising of the single TE conductor operating in two periodical steady state modes: the switching periodical mode (P-mode) when hot and cold ends of the TE conductor are periodically instantly reversed and the continuous sinusoidal mode (S-mode) when the temperature of TE conductor edges varies continuously according to sine wave. Power generation and cooling regimes of rotating (TE) device in the periodic steady state was studied analytically. The efficiency and the cooling temperature of rotating TE device was found to depend not only on the dimensionless TE figure of merit, but also upon an additional dimensionless parameter comprising of the rotation period, the size and the thermal diffusivity of the TE conductor. The proposed analytical method can be generalized to even more complex timing modes and allows to solve the optimization problem for TE device parameters. We investigated whether it is possible to achieve better performance for rotating TE device comparing to conventional stationary steady state, the S-mode was shown to demonstrate deeper cooling at certain times.

----------------------------

Corresponding author: Dr. Igor Bezsudnov, e-mail: biv@akuan.ru

# 1. Introduction

The main way to improve the efficiency of thermoelectric (TE) devices – power generators, coolers etc. is to increase the dimensionless figure of merit of TE materials - $ZT = \alpha^2 \sigma T / \kappa$, where $\alpha$ - the thermopower or the Seebeck coefficient, $\sigma$ - the electrical conductivity, $T$ - the absolute temperature, $\kappa$ - the thermal conductivity.

Unlike of the superconductivity, where new materials with high transition temperatures to the superconducting state have been invented, the progress in $ZT$ improvement of TE materials is quite disappointing. Thus, for example, at room temperature ($T = 300°K$) from 1950 to the present time the figure of merit rose from $ZT \sim 1$ to $ZT \sim 1.2 \div 1.3$ only [1-5]. Moreover, today there is no commercially available TE materials with $ZT \sim 1.3$. Indeed, for common appliances use, for example, in the household or industrial refrigeration, TE materials with the figure of merit $ZT \geq 2.0$ [6-8] are required. There were expectations that the success can be achieved using tunneling and other quantum effects in nanostructured TE materials [5, 9-13]. However, there is no significant progress so far.

The parameters of TE device in the stationary steady state depend only on the figure of merit $ZT$ [14]. The higher $ZT$, the lower cooling temperature can be reached.

In transient modes, the efficiency of TE device is affected by many other parameters such as the temperature diffusivity, the current pulse duration in a pulsed mode [6, 24-29], the relaxation time of thermal processes etc. Such transient modes are constantly attract the attention of researchers [5, 15-29] because such modes have advantages over the stationary steady state. For example, at certain times in a pulsed cooling mode [24-29] deeper cooling can be reached. Optimization of transient mode parameters allows to improve the operation of TE device as compared with the stationary steady state even the same TE materials are used.

Qualitatively, the improved performance of TE devices in the transient mode is possible due to the fact that the relaxation time of electrical processes is negligible compared to the relaxation time of thermal processes [14]. When the current flows through TE device in the cooling regime in the stationary steady state, the Peltier heat removed from the cold junction and the Joule heat generated in the TE conductor are balanced. Increased current and, consequently, increased Joule heat would make the TE device inoperative. In the transient state, due to the relaxation times difference, the heat balance is uncompensated. Higher current passed through the TE device for a short time delivers additional cooling. Optimization of length and shape of the current pulses can give deeper cooling on limited time intervals [27] or the cooling of small objects in a shorter time [28, 29].

The pulsed cooling [15-17] consists of two major phases. The first phase is highly transient one implementing fast and deep cooling, the second phase is the relaxation, in this phase, as a rule, the TE device is out of use.

This paper studies TE devices operating in the periodic steady state mode. Unlike the pulsed cooling, the TE device in the periodic steady state mode operates continuously. The basic question considered here, whether it is possible in this periodic steady state mode to achieve better performance relative to the stationary steady state mode, at least better at certain times. In this study we omit particular technical details such as the contact resistance of the plates, the lateral heat transfer, parameters of the cooled object etc.

The proposed TE devices consist of a single TE conductor with the constant cross section made of thermoelectric material and the role of second conductor plays body of the TE device, which is an ordinary metal conductor.

We consider two types of periodic steady state modes for proposed TE devices: the switching periodical mode (P-mode) when hot and cold ends of TE conductor are periodically instantly reversed and the continuous sinusoidal mode (S-mode) when temperature of TE conductor edges varies continuously according to sine wave.

For periodic steady state modes, along with $ZT$ we found a new dimensionless parameter that is the combination of the period of temperature change, the TE conductor size and its temperature diffusivity. The optimal value of above parameter was calculated.

In the next section TE devices in P- and S-modes are schematically described. Following sections contain analytical calculations and results for P-mode in the power generation and cooling regimes, and for the S-mode cooling regime. The last section presents discussion and conclusions.

## 2. Model of TE device in a periodical steady state.

The TE device operating in the switching periodical mode (P-mode) is shown schematically in Fig. 1.a. The TE conductor turns periodically in the plane of the figure and its hot and cold ends (junctions) are instantly swapped.

The TE device operating in the continuous sinusoidal mode (S-mode) is presented schematically in Fig 1.b. Let the TE conductor rotates in the hole of the orifice plate with linear temperature distribution from up to down (see Fig. 1.b), consequently at the ends (junctions) of the rotating TE conductor (see Fig. 1.b) the temperature varies continuously by sine wave.

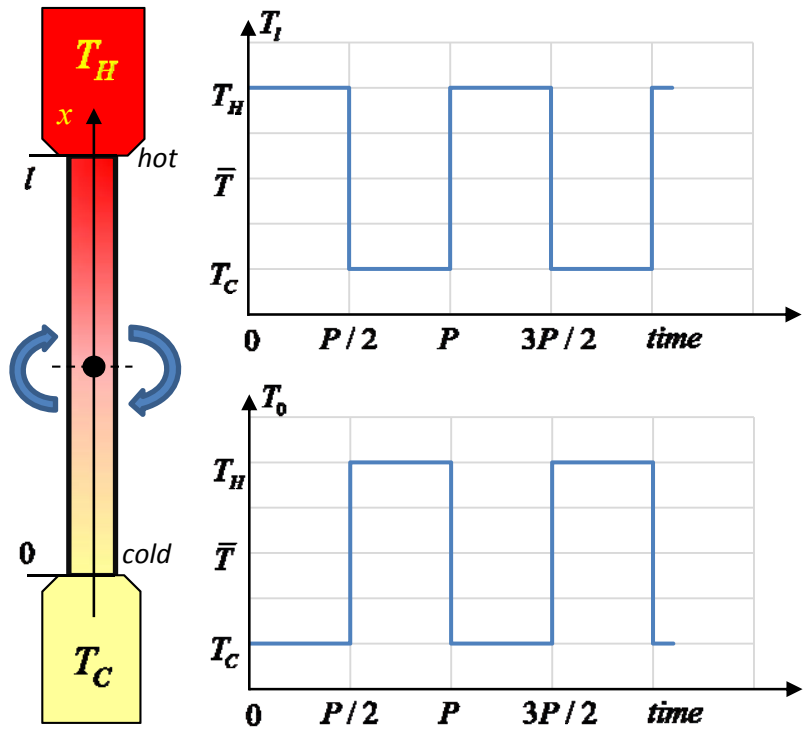
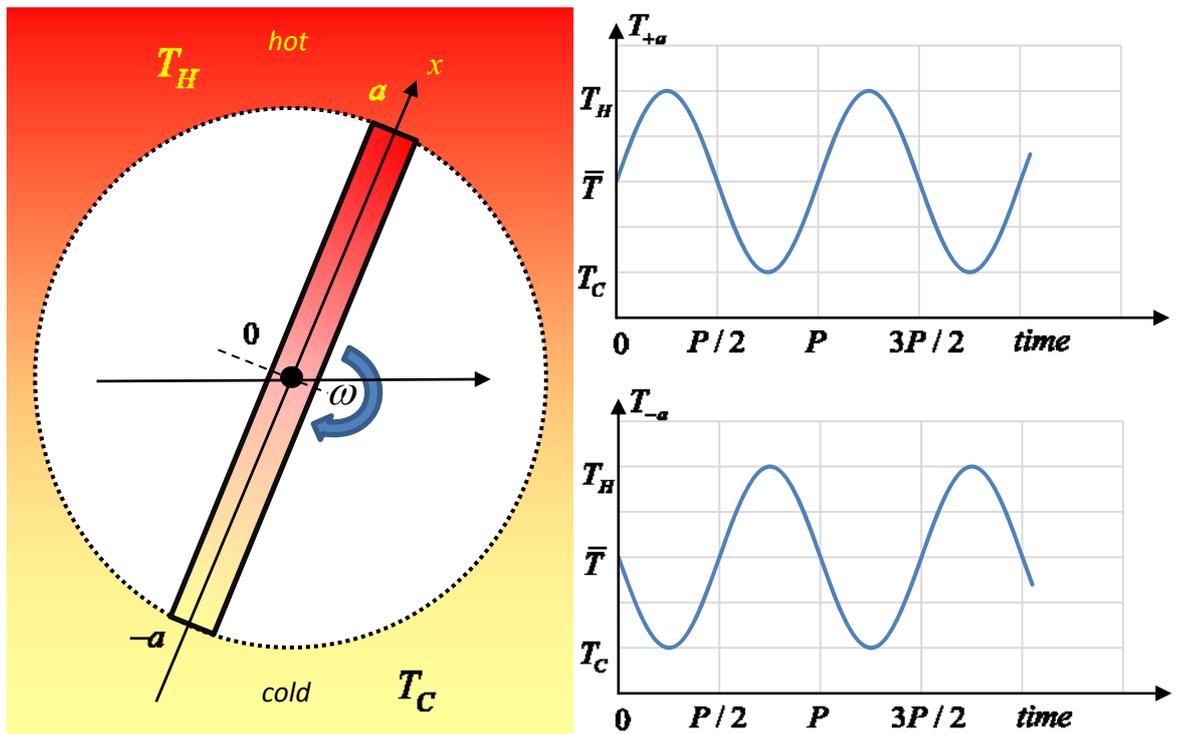

Fig.1. Schematic sketch of proposed TE devices operating in a) the switching periodical mode (P-mode) and b) the continuous sinusoidal mode (S-mode). (color online)

The TE device (Fig. 1a, b) consists of a single TE conductor with the constant cross section $S$ of the length $l = 2a$. Other parts of TE device do not have TE properties. The period of rotation $P$ is fixed.

Performing further analytical calculations for both P- and S-modes we assume for convenience that the TE conductor is fixed in plane but the temperature at its ends (junctions) varies according to the periodic law specific for each mode.

The heat conduction equation for the TE conductor in TE devices has the standard form [14]

$$c_v \rho_0 \frac{\partial T}{\partial t} = \kappa \frac{\partial^2 T}{\partial x^2} + \rho j^2, \qquad (1)$$

where $t$ - the time, $x$ - the coordinate along TE conductor, $T(x,t)$ – the temperature of TE conductor, $j(t)$ – the current density in TE conductor, $\rho = 1/\sigma$ - the specific resistivity, $\kappa$ – the thermal conductivity, $c_v$ – the specific heat, $\rho_0$ – the bulk density, also we denote $\chi = \kappa/c_v \rho_0$ – the thermal diffusivity.

The P-mode (Fig. 1.a) the boundary conditions are as follows

$$\begin{aligned} T(x,t)\big|_{x=0} &= \overline{T} - T_0 \theta(t) \\ T(x,t)\big|_{x=l} &= \overline{T} + T_0 \theta(t) \end{aligned}, \qquad (2)$$

where $\overline{T}$ – the external mean temperature, $T_0$ – the amplitude of variation of the external temperature, the function $\theta(t)$ is set to $-1$ on the even half-periods and on the odd ones is equal to $+1$

$$\theta(t) = \begin{cases} +1, & nP < t < (n+1/2)P \\ -1, & (n+1/2)P < t < (n+1)P \end{cases}. \qquad (3)$$

The S-mode (Fig 1.b) corresponds to the case when the temperature of the ends (junctions) of TE conductor varies continuously according to sine wave, therefore the boundary conditions in the S-mode are

$$T(x,t)\big|_{x=\pm a} = \overline{T} \pm T_0 \sin(\omega t), \qquad (4)$$

where $\omega = P/2\pi$ - the angular frequency of temperature change, $\overline{T}$ and $T_0$ have the same meanings as in the P-mode.

Thus, during the period the TE conductor in P- and S-modes has the maximum temperature at the hot end (junction) $T_H = \overline{T} + T_0$ and minimal at cold end (junction) $T_C = \overline{T} - T_0$.

The current that flows through the TE conductor in the cooling regime is set to

$$j = j_0 \theta(t), \qquad \text{(P– mode)}$$
$$j = j_0 \sin\left(\frac{2\pi}{P}t\right). \qquad \text{(S– mode)} \tag{5}$$

In the power generation regime the TE conductor current is calculated according to Seebeck's law $j \sim \alpha \Delta T$, where α – the thermo power or the Seebeck coefficient, we assume it to be temperature independent

$$\Delta T = 2T_0, \qquad \text{(P– mode)}$$
$$\Delta T = 2T_0 \sin\left(\frac{2\pi}{P}t\right). \qquad \text{(S– mode)} \tag{6}$$

where $2T_0 = \Delta T$ - the maximum temperature difference between hot and cold ends (junctions).

The equation (1) with the boundary conditions (2) or (4) and the relations for the TE conductor current (5) or (6) in the periodic steady state are solved analytically in following sections.

## 3. Temperature distribution and heat fluxes in the switching periodical mode (P-mode)

### 3.1 The temperature distribution in P-mode

The solution of the equation (1) with the boundary conditions (2) for the P-mode was analytically calculated using the method described in [30, Chapter 15].

First, we represent $T(x,t)$ in the following form

$$T(x,t) = \overline{T} + \frac{\rho j^2}{2\kappa} x(l-x) + \tilde{T}(x,t). \tag{7}$$

Then (1) gives the equation for $\tilde{T}(x,t)$

$$\frac{\partial \tilde{T}}{\partial t} = \chi \frac{\partial^2 \tilde{T}}{\partial x} \tag{8}$$

and the boundary conditions (2) become

$$\tilde{T}(x,t)\Big|_{x=0} = -T_0 \theta(t)$$
$$\tilde{T}(x,t)\Big|_{x=l} = +T_0 \theta(t) \tag{9}$$

Next, according to [30] we write $\tilde{T}(x,t)$ in the form of series

$$\tilde{T}(x,t) = \sum_{k=1}^{\infty} T_k(t) \sin\frac{k\pi}{l}x. \tag{10}$$

Substituting $\tilde{T}(x,t)$ (10) in the equation (8), integrating by parts twice and using the boundary conditions (9) we obtain the following relation for $T_k(t)$

$$\frac{dT_k}{dt} + \chi\left(\frac{k\pi}{l}\right)^2 T_k = -\chi\frac{2\pi k}{l^2} T_0 \theta(t)\left[1+(-1)^k\right] \tag{11}$$

The solution of the ordinary differential equation (11) is as follows [31-33]

$$T_k(t) = T_0 e^{-A_k t} - \chi\frac{2\pi k}{l^2}\left[1+(-1)^k\right] e^{-A_k t} \int_0^t \theta(t) e^{A_k t} dt, \tag{12}$$

where

$$A_k = \chi\left(\frac{k\pi}{l}\right)^2. \tag{13}$$

At longer times when the periodic steady state is reached, the first transient term have to disappear.

Let $t = mP + \tau$ where $0 < \tau < P/2$ and $m \gg 1$, i.e. the time $\tau$ is measured from the beginning of the period and at that time the left junction ($x=0$) is cold one and right one ($x=l$) is hot one (see. Fig. 1.a). Then, according to (3), the integral in (12) is divided into three terms, which represent the sum of odd ($\theta(t) = +1$) and even ($\theta(t) = -1$) half-periods, and the third term, which depends on $\tau$

$$\int_0^t \theta(t) e^{A_k t} dt = \sum_{n=0}^m \int_{nP}^{\left(n+\frac{1}{2}\right)P} e^{A_k t} dt + -\sum_{n=0}^m \int_{\left(n+\frac{1}{2}\right)P}^{(n+1)P} e^{A_k t} dt + \int_{(m+1)P}^{(m+1)P+\tau} e^{A_k t} dt. \tag{14}$$

Calculating integrals in the first and second terms of (14) and considering $e^{-mA_k} \ll 1$ at $m \gg 1$, we obtain geometric progressions. The sums of above progressions we use in (12) to get the final formulae

$$T_k(\tau) = -T_0 \frac{2}{k\pi}\left[1+(-1)^k\right] + 2T_0 \frac{2}{k\pi}\left[1+(-1)^k\right] \frac{e^{-A_k t}}{1+e^{A_k \frac{P}{2}}} \tag{15}$$

Substituting now (15) in (1) and considering that

$$\sum_{k=1}^\infty \frac{1}{k} \sin\frac{k\pi}{l} x = \frac{\pi}{2}\left(1-\frac{x}{l}\right) \tag{16}$$

we obtain

$$\tilde{T}(x,\tau) = -T_0\left(1-2\frac{x}{l}\right) + T_0 \sum_{k=1}^\infty N_k e^{-A_k \tau} \sin\frac{k\pi}{l} x, \tag{17}$$

where

$$N_k = \frac{4}{k\pi} \cdot \frac{1+(-1)^k}{1-e^{A_k \frac{P}{2}}} \qquad (18)$$

Finally, the solution (1) with the boundary conditions (3) for the P-mode has the form

$$T(x,\tau) = \bar{T} + \frac{\rho j^2}{2\kappa} x(l-x) - T_0\left(1 - 2\frac{x}{l}\right) + T_0 \sum_{k=1}^{\infty} N_k e^{-A_k \tau} \sin\frac{k\pi}{l} x, \qquad (19)$$

where $\tau$ belongs to $[0..P/2]$.

Fig.2 shows the temperature distribution along the TE conductor in the P-mode at different times.

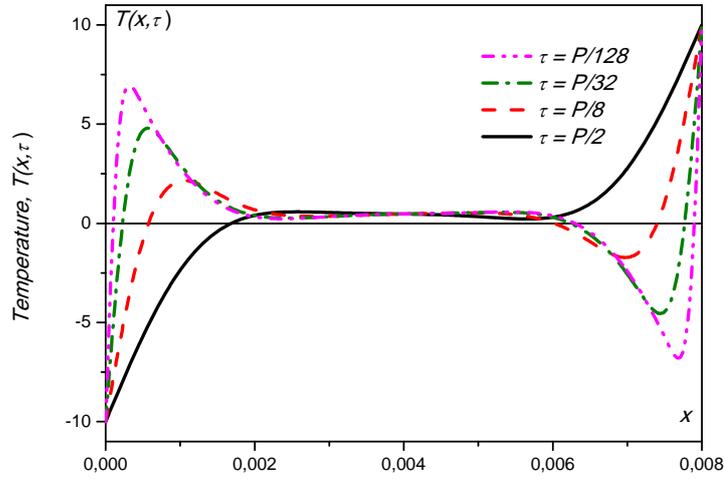

*Fig.2. Temperature distribution of TE conductor in P-mode at different times measured from the begin of the period: $P/128...P/2$. External temperatures: $\bar{T} = 0°C$, $T_0 = 10°C$, TE conductor length $l = 0.08m$, TE material parameters $\kappa = 1.7 W/mK$, $\chi = 1.2 \cdot 10^{-6} m^2/s$, current density $j = 0.1 \cdot 10^6 A/m^2$ and rotation period $P = 1s$. (color online)*

At $t = P, 2P, 3P,...$ the temperature of TE conductor edges is changing instantly. Then after this leap the TE conductor starts warming up, it is clearly seen from Fig. 2. but there is still a part of TE conductor with the temperature $T(x,\tau) < \bar{T}$. Eventually the size of this part decreases and its temperature increases.

Selected TE conductor and operational parameters show the case when temperature in the middle of the TE conductor is stable i.e. heat waves do not enter deeply in the TE conductor. It is similar to permafrost, when a periodic variation of the temperature on Earth's surface does not affect the temperature at a certain depth.

It should be noted also that the temperature in the middle of TE conductor is slightly higher than $\bar{T}$ because of the emitted Joule heat. When no current the temperature in the center of TE conductor, of course, would be equal to $\bar{T}$.

### 3.2 Power generation efficiency in P-mode

The power generation efficiency $\eta$ of TE device in the P-mode depends on the heat flux coming into the hot junction and coming out of the cold junction of the TE conductor (at $x = 0$ and $x = l$ respectively) [14]. The heat flux density $q_x$ is the sum of the flux densities created by the temperature distribution $\sim \partial T / \partial x$ and the Peltier heat flux $\Pi \mathbf{j}$, where $\Pi = \alpha T$ is the Peltier coefficient (we assume that the thermo power or the Seebeck coefficient α is temperature independent).

$$q_k = -\kappa \frac{\partial T}{\partial x} - \alpha T j \ . \tag{20}$$

We find the heat flux at cold - $\dot{Q}_C$ and hot - $\dot{Q}_H$ junctions using (19)

$$\dot{Q}_C = q_x\big|_{x=0} = -\frac{\rho J^2}{2}\frac{l}{S} - \kappa \frac{2T_0}{l}S - \kappa T_0 S \sum_{k=1}^{\infty} \frac{k\pi}{l} N_k e^{-A_k \tau} + \alpha T_c J \ , \tag{21}$$

$$\dot{Q}_H = q_x\big|_{x=l} = -\frac{\rho J^2}{2}\frac{l}{S} - \kappa \frac{2T_0}{l}S - \kappa T_0 S \sum_{k=1}^{\infty} (-1)^k \frac{k\pi}{l} N_k e^{-A_k \tau} + \alpha T_H J \ . \tag{22}$$

Here $S$ - the cross section of TE conductor, $J$ - the current flows through TE conductor. $J$ is governs by the emf $\varepsilon$ generated according to the Seebeck effect $\varepsilon = \alpha(T_H - T_c)$ and by connected in series the TE conductor resistance $r = \rho l / S$ and the load resistance $R$:

$$J = \alpha \frac{T_H - T_C}{r + R} = \alpha \frac{2T}{r(1+\Omega)}, \quad \Omega = \frac{r}{R} \ . \tag{23}$$

The signs of Peltier heat terms in (21) and (22) are selected in the way that the flux positive direction is from the hot to the cold junction i.e. the heat flux coming to the hot junction ($x = l$) and coming out of the cold ($x = 0$) junction is set to be positive.

The heat coming out of the hot $Q_H$ and coming in the cold $Q_C$ junctions depends on time, therefore to get the efficiency $\eta$ we have to integrate $Q_H$ and $Q_C$ for a certain time, such time for the P-mode is $P/2$ - half of the period:

$$Q_H = \int_0^{P/2} \dot{Q}_H d\tau, \qquad Q_C = \int_0^{P/2} \dot{Q}_C d\tau \ . \tag{24}$$

Substituting in (24) the expressions for $\dot{Q}_H$ (21) and $\dot{Q}_C$ (22) we find

$$\frac{1}{P/2}Q_C = \frac{1}{2}rJ^2 + \frac{2T_0}{l}S\kappa_e + \alpha T_c J$$
$$\frac{1}{P/2}Q_H = -\frac{1}{2}rJ^2 + \frac{2T_0}{l}S\kappa_e + \alpha T_H J \quad , \quad (25)$$

where the renormalized thermal conductivity is

$$\kappa_e = \kappa\left[1 + \frac{4\mu_0^2}{\pi}\sum_{k=1}^{\infty}\frac{1+(-1)^k}{(k\pi)^2}th\left(A_k\frac{P}{4}\right)\right] \quad (26)$$

and

$$\mu_0^2 = l^2\frac{1}{P\chi}. \quad (27)$$

Comparing the relations (25) and (26) for the P-mode and the formulae for the stationary steady state [14] one can realize that they differs only in the thermal conductivity value. The efficiency $\eta$ in the stationary steady state depends only on the thermal conductivity - $\kappa$, but in the P-mode it depends upon the renormalized thermal conductivity $\kappa_e$ (26), which is a complex parameter proportional not only to the TE conductor thermal conductivity - $\kappa$, but also to the length of TE conductor, the switching period $P$ and the thermal diffusivity - $\chi$.

Therefore the corresponding expression for the efficiency $\eta = (Q_H - Q_C)/Q_H$ for the P-mode is similar to the stationary steady state but it uses the renormalized thermal conductivity $\kappa_e$ (26). Appropriate calculations can be found for instance in [14] and below is the final expression

$$\eta = \frac{Z_e \Delta T \Omega}{(1+\Omega)^2\left[1 + \frac{Z_e T_H}{1+\Omega} - \frac{1}{2}\frac{Z_e \Delta T}{(1+\Omega)^2}\right]}, \quad (28)$$

where $Z_e = \sigma\alpha^2/\kappa_e$ - the renormalized using (26) dimensionless TE figure of merit.

As in the stationary steady state [14], the maximum efficiency $\eta$ in the P-mode is achieved at the optimal ratio $\Omega_{opt} = R/r = \sqrt{1+Z_e\overline{T}}$. Using $\Omega_{opt}$ in (28) we find the value of maximal efficiency $\eta_{max}$ for the P-mode that depends only on $T_H, T_C$ and $Z_e$

$$\eta_{max} = \frac{\Delta T}{T_H}\frac{\sqrt{1+Z_e\overline{T}}-1}{\sqrt{1+Z_e\overline{T}}-\frac{T_C}{T_H}}. \quad (29)$$

The maximum efficiency $\eta_{max}$ is a monotonically increasing function of $Z_e$, thus higher $Z_e$ and accordingly lower $\kappa_e$ give the better $\eta_{max}$ value.

The renormalized thermal conductivity $\kappa_e$ in the P-mode (26) is always greater than $\kappa$, $\kappa_e > \kappa$ thus the efficiency in the P-mode (28) is always less than the efficiency in the stationary steady state. In the case when $\chi P\pi^2/4l > 3$ the hyperbolic tangent in (26) is nearly one and considering $\sum_{k=1}^{\infty} \frac{\left[1+(-1)^k\right]}{(k\pi)^2} = \frac{1}{12}$ we get the approximated expression for the renormalized thermal conductivity

$$\kappa_e \approx \kappa\left(1 + \frac{1}{3}\mu_0^2\right) \qquad (30)$$

To have the efficiency of TE device in the P-mode as high as possible we need $\kappa_e \to \kappa$ or $\mu_0^2 \to 0$ (30). The latter means higher $\chi$ values or shorter lengths $l$ of the TE conductor. In other words, for half of the period the TE conductor to be warmed almost as the TE conductor in the stationary steady state.

### 3.3 Cooling in P-mode

Calculations for the cooling regime in the P-mode are similar to those for the efficiency, should only take into account that the current is determined by (5) $J = J_0 \theta(t)$ but not by the Seebeck effect. The optimal current $J_{opt}$ minimize the cooling temperature or maximize the coefficient of performance $K$.

The heat fluxes $\dot{Q}_C$ and $\dot{Q}_H$ in the P-mode cooling regime differ from (21) and (22) only by the signs of the Peltier heat term because the current in the TE conductor flows in inverse direction than in the power generation regime.

$$\left.\begin{array}{l} \dfrac{1}{P/2}\dot{Q}_C = \dfrac{1}{2}rJ^2 + \dfrac{2T_0}{l}S\kappa_e - \alpha T_C J \\[6pt] \dfrac{1}{P/2}\dot{Q}_H = -\dfrac{1}{2}rJ^2 + \dfrac{2T_0}{l}S\kappa_e - \alpha T_H J \end{array}\right\}. \qquad (31)$$

In the cooling regime we have to follow variations of the effective (normalized) thermo conductivity $\kappa_e(\tau)$ to find the time when the lowest possible cooling temperature can be reached. $\kappa_e(\tau)$ depend on the time $\tau$ as follows

$$\kappa_e(\tau) = \kappa\left[1 + \frac{1}{2}\sum_{k=1}^{\infty} k\pi N_k e^{-A_k \tau}\right]. \qquad (32)$$

It should be noted that $\kappa_e(\tau)$ in contrast to $\kappa_e$ (26) do not use the factor $\mu_0^2$ (27).

Next, as in the stationary steady state, the condition $\partial \dot{Q}_C / \partial J = 0$ for the current gives

$$J_{opt} = \frac{\alpha T_C}{\rho l}S = \frac{\alpha T_C}{r}. \qquad (33)$$

Note that in contrast to the stationary steady state, the thermo conductivity $\kappa_e(\tau)$ (32) depends on the time $\tau$, however, it does not affect the value of the optimal current $J_{opt}$.

The expression for the heat flux at the cold junction at the optimum current is equal to

$$\frac{1}{S}\dot{Q}_C = -\frac{1}{2}\frac{\alpha^2 T_c^2}{\rho l} + \frac{2T_0}{l}\kappa_e(\tau) \qquad (34)$$

In the stationary steady state, the minimum cooling temperature $T_c^{\min}$ is found from the condition $\dot{Q}_c = 0$. In the P-mode the condition $\dot{Q}_c = 0$ is only possible at certain times. Assuming $\dot{Q}_c = 0$ (34) we get

$$T_c^{\min} = 2\overline{T}\frac{\sqrt{1+Z_e(\tau)\overline{T}}-1}{Z_e(\tau)\overline{T}}, \qquad (35)$$

where the renormalized figure of merit $Z_e(\tau) = \sigma\alpha^2 / \kappa_e(\tau)$ depends on the time $\tau$.

The only difference between $T_c^{\min}(\tau)$ (35) and the expression for $T_c^{\min}$ in the stationary steady state is the value of figure of merit. The $T_c^{\min}(\tau)$ (35) uses the renormalized thermal conductivity $\kappa_e(\tau)$ which depends on the time $\tau$ and allows to perform the optimization of the cooling temperature.

The $Z_e(\tau)$ maximum value, i.e. the minimum cooling temperature $T_c^{\min}$ corresponds to the minimum $T_c^{\min}$. As it follows from (32), $\kappa_e(\tau)$ has the minimum at $\tau = P/2$, however, even in this case $\kappa_e(\tau = P/2) > \kappa$ i.e. similar to the P-mode power generation regime.

Due to the fact that $\kappa_e(\tau) > \kappa$, the coefficient of performance $K = \dot{Q}_C/(\dot{Q}_H - \dot{Q}_C)$ and the maximum cooling capacity $Q_c^{\max}$ are less than those in the stationary steady state, although at certain values of the thermo conductivity, the switching period and other TE conductor parameters are close to it.

The above conclusions apply only to the P-mode. Further, we show that the TE device operating S-mode can demonstrate better performance.

## 4. Temperature distribution and heat flux in S-mode
### 4.1 The temperature distribution in S-mode

The solution of the equation (1) with the boundary conditions (4) for the S-mode begins from representing $T(x,t)$ in the form

$$T(x,t) = \overline{T} - j_0^2 F\left(\sin 2\omega t - \mu^2\frac{a^2-x^2}{a^2}\right) + \tilde{T}(x,t), \qquad (36)$$

where

$$\mu^2 = a^2 \frac{\omega}{\chi}, \quad F = \frac{\rho}{4c_v \rho_0 \omega}. \tag{37}$$

Then (1) gives the equation for $\tilde{T}(x,t)$

$$\frac{\partial \tilde{T}}{\partial t} = \chi \frac{\partial^2 \tilde{T}}{\partial x^2} \tag{38}$$

and the boundary conditions (4) becomes

$$\tilde{T}(x,t)\big|_{x=\pm a} = \pm T_0 \sin(\omega t) + j_0^2 \, F \sin(2\omega t). \tag{39}$$

The solution of (38) with the boundary conditions can be found in the form

$$\tilde{T}(x,t) = T_0 \big(S(x)\sin \omega t + C(x)\cos \omega t\big) + j_0^2 \, F\big(\tilde{S}(x)\sin 2\omega t + \tilde{C}(x)\cos 2\omega t\big), \tag{40}$$

where

$$\begin{cases} S(x) = S_{cs} ch(\frac{\mu}{\sqrt{2}a}x)\sin(\frac{\mu}{\sqrt{2}a}x) + S_{sc} sh(\frac{\mu}{\sqrt{2}a}x)\cos(\frac{\mu}{\sqrt{2}a}x), \\ C(x) = S_{sc} ch(\frac{\mu}{\sqrt{2}a}x)\sin(\frac{\mu}{\sqrt{2}a}x) - S_{cs} sh(\frac{\mu}{\sqrt{2}a}x)\cos(\frac{\mu}{\sqrt{2}a}x), \\ \tilde{S}(x) = \tilde{S}_{cc} ch(\frac{\mu}{a}x)\cos(\frac{\mu}{a}x) + \tilde{S}_{ss} sh(\frac{\mu}{a}x)\sin(\frac{\mu}{a}x), \\ \tilde{C}(x) = -\tilde{S}_{ss} ch(\frac{\mu}{a}x)\cos(\frac{\mu}{a}x) + \tilde{S}_{cc} sh(\frac{\mu}{a}x)\sin(\frac{\mu}{a}x). \end{cases} \tag{41}$$

The boundary conditions (39) allow to find coefficients in (41):

$$S_{cs} = \frac{ch(\mu/\sqrt{2})\sin(\mu/\sqrt{2})}{sh^2(\mu/\sqrt{2}) + \sin^2(\mu/\sqrt{2})}, \quad S_{sc} = \frac{sh(\mu/\sqrt{2})\cos(\mu/\sqrt{2})}{sh^2(\mu/\sqrt{2}) + \sin^2(\mu/\sqrt{2})},$$

$$\tilde{S}_{ss} = \frac{sh(\mu)\sin(\mu)}{ch^2(\mu) - \sin^2(\mu)}, \quad \tilde{S}_{cc} = \frac{ch(\mu)\cos(\mu)}{ch^2(\mu) - \sin^2(\mu)}. \tag{42}$$

Therefore the solution of (1) with the boundary conditions (4) can be written as

$$T(x,t) = \overline{T} + T_0 \big(S(x)\sin \omega t + C(x)\cos \omega t\big) + \\ + j_0^2 \, F\left(\sin 2\omega t + \mu^2 \frac{a^2 - x^2}{a^2} + \tilde{S}(x)\sin 2\omega t + \tilde{C}(x)\cos 2\omega t\right). \tag{43}$$

The solution $T(x,t)$ (43) includes both the terms proportional to $\sin \omega t, \cos \omega t$ and those proportional to $\sin 2\omega t, \cos 2\omega t$ - double the frequency of the temperature change at the TE conductor ends (junctions). Terms in (43) with $\sin \omega t$ and $\cos \omega t$ owe their origin from the heat flux generated by the temperature difference at the TE conductor ends ($x = \pm a$), their amplitude is proportional to $T_0$. Such terms describe common temperature waves damping with the distance from TE conductor ends ($x = \pm a$). Parameter $\mu^2$ is a combination of the frequency, the length and the thermal diffusivity of TE conductor which similar to $\mu_0^2$ in the P-mode (27).

The double frequency terms $\sin 2\omega t$, $\cos 2\omega t$ in (43) are proportional to the square of the amplitude of the current density amplitude $j_0^2$. Those terms describe the heat flux, born by the heterogeneity of the temperature distribution due to the Joule heat.

On Fig 3. we present the temperature distribution in the TE conductor which has the same parameters as on Fig.2 above but used in the S-mode.

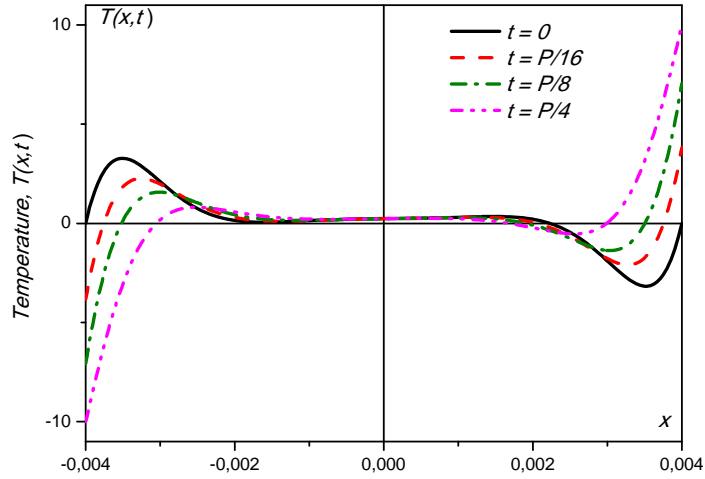

Fig.3. Temperature distribution of TE conductor in S-mode at different times measured from the begin of the period: $0...P/16$. External temperatures: $\bar{T} = 0°C$, $T_0 = 10°C$, TE conductor size $a = 0.04m$, TE material parameters $\kappa = 1.7W/mK$, $\chi = 1.2 \cdot 10^{-6} m^2/s$, current density amplitude $j = 0.1 \cdot 10^6 A/m^2$ and rotation period $P = 1s$. (color online)

Fig. 3 shows that the temperature distribution of TE conductor in the S - mode is similar to that for P-mode. But the temperature at the TE conductor edges in the S-mode is changed continuously by sine wave in the range $\bar{T} \pm T_0$. In the center of the TE conductor $T(0,t) > \bar{T}$ because of Joule heat emission.

## 4.2 Cooling in S-mode

Let during the first half-period the TE conductor has the lower end (junction) temperature (see Fig.1.b) colder than upper one, i.e. the lower junction is cold one.

Therefore the heat flux at the cold junction $(x = -a)$ is

$$Q_c = -q_x S\big|_{x=-a} = S\kappa \frac{\partial T}{\partial x}\bigg|_{x=-a} - \alpha T_c(t) j(t) S, \tag{44}$$

where the second term is the Peltier heat flux, $S$ - the cross section of TE conductor, $T_c(t) = T(x = -a, t)$.

Substituting in (44) the expression for the temperature distribution from (43) at $x = -a$ we obtain

$$\frac{Q_c}{S}\bigg|_{x=-a} = \kappa T_0 \left( S'(-a) \sin \omega t + C'(-a) \cos \omega t \right) +$$
$$+ \kappa j_0^2 \, F \left( -\frac{2\mu^2}{a} + \tilde{S}'(-a) \sin 2\omega t + \tilde{C}'(-a) \cos 2\omega t \right) - \alpha T_c(t) j(t). \tag{45}$$

At the cold junction acc. to (4) the temperature will be minimal at $\omega t = \pi/2$. We denote it as $T_c$. At that moment the heat flux at the cold junction is

$$\frac{Q_c}{S}\bigg|_{\substack{x=-a \\ \omega t=\pi/2}} = \kappa T_0 S'(-a) - \kappa j_0^2 \, F \left( 2\mu^2/a - \tilde{C}'(-a) \right) - \alpha T_c j_0. \tag{46}$$

To determine the possible lowest cooling temperature of the cold junction we have to find minimum of the heat flux at the cold junction i.e. $Q_c/S \big|_{\substack{x=-a \\ \omega t=\pi/2}}$ as the function of current density amplitude $j_0$, then to find the optimal current $j_0^{opt}$ we use the condition $\partial Q_c / \partial j_0 = 0$. Further, using $j_0^{opt}$ we calculate the heat flux at the cold junction (46) and finally obtain the minimal cooling temperature $T_c^{min}$.

The optimal current value $j_0^{opt}$ is

$$j_0^{opt} = \frac{\alpha T_c}{2 F \kappa \left( 2\mu^2/a - \tilde{C}'(-a) \right)}. \tag{47}$$

At the minimal temperature $T_c^{min}$ the heat flux $Q_c$ of the cold junction at the current density amplitude $j_0^{opt}$ to be equal to zero

$$\frac{Q_c}{S}\bigg|_{\substack{x=-a \\ \omega t=\pi/2}} = \kappa T_0 S'(-a) - \frac{(\alpha T_c)^2}{4 F \kappa} \frac{1}{\left( 2\mu^2/a - \tilde{C}'(-a) \right)} = 0. \tag{48}$$

Using $T_0 = \overline{T} - T_c$ for (48) we find

$$T_c^2 = 2\overline{T}(\overline{T} - T)_c \left[ \frac{2 F \kappa^2}{\overline{T} \alpha^2 a^2} \left( 2\mu^2 - a\tilde{C}'(-a) \right) a S'(-a) \right]. \tag{49}$$

Let us denote the term in square brackets in (49) $\beta(\mu, Z\overline{T})/Z\overline{T}$ then using $2F\kappa^2/\overline{T}\alpha^2 a^2 = 1/2Z\overline{T}\mu^2$ we write the following formulae for the dimensionless parameter $\beta(\mu)$

$$\beta(\mu) = \frac{\mu}{2\sqrt{2}}\left(1 + \frac{1}{2\mu}\frac{sh(\mu)ch(\mu) + \cos(\mu)\sin(\mu)}{ch^2(\mu) - \sin^2(\mu)}\right) \times \\ \times \frac{sh(\mu/\sqrt{2})ch(\mu/\sqrt{2}) + \sin(\mu/\sqrt{2})\cos(\mu/\sqrt{2})}{sh^2(\mu/\sqrt{2}) + \sin^2(\mu/\sqrt{2})} \quad (50)$$

and the equation (49) is rewritten in the form

$$T_c^2 + \frac{\beta(\mu)}{Z\overline{T}}\overline{T}T_c - \frac{\beta(\mu)}{Z\overline{T}}\overline{T}^2 = 0 \quad (51)$$

To calculate $T_c^{min}$ we have to solve the quadratic equation (51), the positive root of (51) gives $T_c^{min}$

$$T_c^{min} = \frac{2\overline{T}}{1 + \sqrt{1 + \frac{Z\overline{T}}{\beta(\mu)}}}. \quad (52)$$

Let us compare $T_c^{min}$ in the S-mode to the minimal cooling temperature in the stationary steady state - $T_c^{st}$ that can be expressed [14] as follows

$$T_c^{st} = \frac{T_H}{\sqrt{1 + Z\overline{T}}}, \quad (53)$$

or using $T_H = 2\overline{T} - T_c$

$$T_c^{st} = \frac{2\overline{T}}{1 + \sqrt{1 + Z\overline{T}}}. \quad (54)$$

Relations (52) and (54) have the similar form, that allows to rewrite the expression for $T_c^{min}$

$$T_c^{min} = \frac{2\overline{T}}{1 + \sqrt{1 + Z^e\overline{T}}}, \quad (55)$$

where $Z^e\overline{T}$ is the renormalized figure of merit

$$Z^e\overline{T} = \frac{Z\overline{T}}{\beta(\mu)}. \quad (56)$$

Thus, the S-mode calculations are also similar to the stationary steady state. As seen from (55) $T_c^{min}$ is a monotonically increasing function of the dimensionless parameter $\beta(\mu)/Z\overline{T}$, at large values $\beta(\mu)/Z\overline{T} \gg 1$ we obtain $T_c^{min} \to \overline{T}$, that means no cooling at large $\beta(\mu)/Z\overline{T}$, i.e. the smaller the value $\beta(\mu)/Z\overline{T}$, the lower $T_c^{min}$.

Also we can state that the higher TE figure of merit $Z\bar{T}$ means better cooling in the S-mode (50) and opposite, when $Z\bar{T} \to 0$, $\beta(\mu)/Z\bar{T} \to \infty$ and the cooling is impossible. Further, note that (50) и (52) show that $T_c^{min}$ depends only on the single dimensionless parameter - $\mu$.

Let examine the ratio of minimal temperatures in the S-mode and in the stationary steady state

$$\frac{T_c^{min}}{T_c^{st}} = \frac{1+\sqrt{1+Z^e \bar{\bar{T}}}}{1+\sqrt{1+Z\bar{T}}} = \frac{1+\sqrt{1+Z\bar{T}/\beta(\mu)}}{1+\sqrt{1+Z\bar{T}}} \quad . \tag{57}$$

When the figure of merit $Z\bar{T}$ is fixed, the ratio (57) depends only on the single parameter $\beta(\mu)$. The function $\beta(\mu)$ has one minimum and it is invariant relative to $Z\bar{T}$. The minimum is at $\mu \approx 1.53$ and accordingly $\beta \approx 0.76 < 1$ i.e. the TE device in the S-mode will deliver deeper cooling at certain times than in the stationary steady state (see Fig.4).

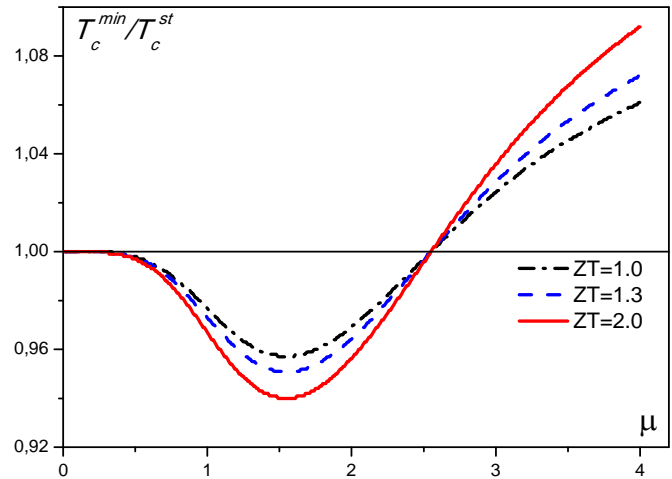

Fig.4. Ratio of minimal temperatures $T_c^{min}/T_c^{st}$ for S-mode and stationary steady state on $\mu$. $1 - Z\bar{T} = 2$, $2 - Z\bar{T} = 1.3$, $3 - Z\bar{T} = 1$. *(color online)*

As an example, for the TE device in the S-mode with the TE conductor made of $Bi_2Te_3$ [34] (the temperature diffusivity $\chi \approx 1.2 \cdot 10^{-6} m^2/s$) we choose the length of the TE conductor $2a = 1.5mm$ and the rotation period $P \approx 1.15 \sec$ that gives optimal $\mu_{min} \approx 1.53$.

For the S-mode at optimal $\beta(\mu) \approx 0.76$ the minimal achievable cooling temperature can be expressed as

$$T_c^{\min} \approx \frac{2\overline{T}}{1+\sqrt{1+1.3Z\overline{T}}} \,. \tag{58}$$

In other words, using the TE material with $ZT=1$ in the S-mode we get cooling as of material with $ZT=1.3$ in the stationary steady state. The TE material with $ZT=1.3$ in S-mode corresponds to $ZT=1.7$ in the stationary steady state.

## 5. Discussion and conclusions

The paper describes two types of proposed TE devices that operates in periodic steady state modes: P-mode – the switching periodical mode and S-mode – the continuous sinusoidal mode.

In common, the efficiency of TE device is related to the rate of entropy production or particularly to the volume integral of the divergence of the entropy flux density $\mathbf{s}=\mathbf{q}/T$, $B=\int_V div(\mathbf{s})dV$. Finally, the efficiency can be written in the form

$$\eta = \eta_c \frac{1}{1+\frac{BT_H}{A}} \,, \tag{59}$$

where $A$ - the work performed by the TE power generator.

Thus, the higher entropy production rate, the lower the efficiency of the TE device in the power generation regime. Transient regimes, having some additional spatial inhomogeneity of the temperature distribution, naturally leads to the additional entropy production $B$ and, as a consequence, the efficiency have to be even lower.

But the relation (59) that binds the efficiency and the entropy production is derived for the stationary steady state. Therefore, to predict what could be the efficiency in transient, pulsed or periodic modes is almost impossible in advance. Generally, the article discuss the possible benefits of TE device usage in transient modes, particularly periodical modes.

The S-mode was shown to demonstrate deeper cooling at certain times, as compared with the stationary steady state.

The proposed method to calculate analytically parameters of TE devices in periodic P- and S-modes for the power generation regime or the cooling regime can be generalized to even more complex timing modes. The analytical solution allows to apply the optimization technique to find optimal TE device parameters.

## Acknowledgments

Authors would like to express their great appreciation to Dr. L.N. Vikhor and Prof. S.Z. Sapozhnikov, for their valuable and constructive suggestions during development of this research work. Also we would like to thank also Prof. I.V. Andrianov for useful discussions.

I.V.Bezsudnov was supported by the Government of the Russian Federation (Grant 074-U01).

# References

1. Terry M. Tritt, Thermoelectric Phenomena, Materials, and Applications, Annual Review of Materials Research, Vol. 41, No. 1. (2011), pp. 433-448
2. G. Jeffrey Snyder, Eric S. Toberer, Complex thermoelectric materials, Nature Materials **7**, 105 - 114 (2008)
3. George S. Nolas, Joe Poon & Mercouri Kanatzidis, Recent developments in bulk thermoelectric materials, Materials Research Society Bulletin, V. 31, 2006, 199-205
4. H. Alam, S. Ramakrishna, A review on the enhancement of figure of merit from bulk to nano-thermoelectric materials, Nano Energy (2013), http://dx.doi.org/10.1016/j.nanoen.2012.10.005 Hilaal Alama, Seeram Ramakrishnab, A review on the enhancement of figure of merit from bulk to nano-thermoelectric materials, Nano EnergyVolume 2, Issue 2, March 2013, Pages 190–212
5. Nolas, G.S., Sharp, J., Goldsmid, J., Thermoelectrics. Basic Principles and New Materials Developments, Springer Series in Materials Science, Vol. 45, 2001, VIII, 293 p.
6. X.F. Zheng, C.X. Liu, Y.Y. Yan, Q. Wang, A review of thermoelectrics research – Recent developments and potentials for sustainable and renewable energy applications, Renewable and Sustainable Energy Reviews 32 (2014) 486–503
7. New Materials for Thermoelectric Applications: Theory and Experiment, NATO Science for Peace and Security Series B: Physics and Biophysics, 2013, XX, 273 p. 60 illus. Zlatic, Veljko, Hewson, Alex (Eds.) Springer
8. Properties and Applications of Thermoelectric Materials The Search for New Materials for Thermoelectric Devices Proceedings of the NATO Advanced Research Workshop on Properties and Application of Thermoelectric Materials, Hvar, Croatia, 21-26 September 2008 Series: NATO Science for Peace and Security Series B: Physics and Biophysics  Zlatic, Veljko, Hewson, Alexander (Eds.)2009
9. Thermoelectric Nanomaterials Materials Design and Applications Series: Springer Series in Materials Science, Vol. 182  Koumoto, Kunihito, Mori, Takao (Eds.) 2013, XIX, 387 p. 190 illus., 72 illus. in color.

10. Weishu Liu, Xiao Yan, Gang Chen, Zhifeng Ren, Recent advances in thermoelectric nanocomposites, Nano Energy (2012) 1, 42-56
11. L. P. Bulat, I. A. Drabkin, V. V. Karatayev V. B. Osvenskii, Yu. N. Parkhomenko, D. A. Pshenay-Severin, A. I. Sorokin, The Influence of Anisotropy and Nanoparticle Size Distribution on the Lattice Thermal Conductivity and the Thermoelectric Figure of Merit of Nanostructured $(Bi,Sb)_2Te_3$, Journal of Electronic Materials, June 2014, Volume 43, Issue 6, pp 2121-2126
12. L. P. Bulat, V. B. Osvenskii, D. A. Pshenai-Severin, Influence of grain size distribution on the lattice thermal conductivity of $Bi_2Te_3$-$Sb_2Te_3$-based nanostructured materials, Physics of the Solid State, December 2013, Volume 55, Issue 12, pp 2442-2449
13. Наша статья
14. Goldsmid, H. Julian, Introduction to Thermoelectricity, Series: Springer Series in Materials Science, Vol. 121, 2009, XVI, 242 p.
15. Stilbans L S and Fedorovich N A 1958 Sov. Phys.—Tech. Phys. **3** 460–3
16. Parrott J E 1960 Solid State Electron. **1** 135–43
17. V. A. Naer, Transient regimes of thermoelectric cooling and heating units, Journal of engineering physics, April 1965, Volume 8, Issue 4, pp 340-344
18. Snyder G J, Fleurial J P, Caillat T, Yang R and Chen G 2002 J. Appl. Phys. **92** 1564–9
19. Hoyos G E, Rao K R and Jerger D 1977 Energy Convers. **17** 45–54
20. Landecker K and Findlay A W 1961 Solid State Electron. **3** 239–60
21. Thonhauser T, Mahan G D, Zikatanov L and Roe J 2004 Appl. Phys. Lett. **85** 3247–9
22. Richard L. Field, Harold A. Blum, Fast transient behavior of thermoelectric coolers with high current pulse and finite cold junction, Energy Conversion Volume 19, Issue 3, 1979, Pages 159–165
23. Q Zhou, Z Bian and A Shakouri, Pulsed cooling of inhomogeneous thermoelectric materials J. Phys. D: Appl. Phys. 40 (2007) 4376–4381
24. G. Jeffrey Snyder, Jean-Pierre Fleurial, Thierry Caillat, Ronggui Yang, Gang Chen, Supercooling of Peltier cooler using a current pulse J. Appl. Phys. August 1, 2002 Volume 92, Issue 3, pp.1564-1569
25. Yu. I. Dudarev, M. Z. Maksimov, Mathematical modeling of the nonstationary operation of thermoelectric current sources. Technical Physics June 1998, Volume 43, Issue 6, pp 737-738

26. Anutosh Chakraborty, Kim Choon Ng, Thermodynamic formulation of temperature–entropy diagram for the transient operation of a pulsed thermoelectric cooler, International Journal of Heat and Mass Transfer, Volume 49, Issues 11–12, June 2006, Pages 1845–1850

27. Ronggui Yanga, Gang Chena, A. Ravi Kumarb, G. Jeffrey Snyderc, Jean-Pierre Fleurialc, Transient cooling of thermoelectric coolers and its applications for microdevices, Energy Conversion and Management, Volume 46, Issues 9–10, June 2005, Pages 1407–1421

28. G. Jeffrey Snyder, Jean-Pierre Fleurial, Thierry Caillat, Ronggui Yang, Gang Chen, Supercooling of Peltier cooler using a current pulse J. Appl. Phys. August 1, 2002 Volume 92, Issue 3, pp.1564-1569

29. Mao, J. N.; Chen, H. X.; Jia, H.; Qian, X. L. The transient behavior of Peltier junctions pulsed with supercooling. Journal of Applied Physics; July 2012

30. Grinberg G.A. Selected problems of mathematical theory of electric and magnetic phenomena. Publ, Ac. of sci. USSR, Moscow, 1948, 727p. (Гринберг Г.А. Избранные вопросы математической теории электрических и магнитных явлений М.-Л.: Издательство Академии наук СССР, 1948. — 727 с.: ил.)

31. Haberman R 1998 Elementary Applied Partial Differential Equations with Fourier Series and Boundary Value Problems (Englewood Cliffs, NJ: Prentice-Hall) pp 215, 343 – тоже какой-то извращенец, но хорошо бы его найти и посотреть.

32. D. Zwillinger, Handbook of Differential Equations (3rd edition), Academic Press, Boston, 1997.

33. A. D. Polyanin and V. F. Zaitsev, Handbook of Exact Solutions for Ordinary Differential Equations (2nd edition), Chapman & Hall/CRC Press, Boca Raton, 2003. ISBN 1-58488-297-2.

34. H.Scherrer, S.Scherrer, Thermoelectric Properties of Bismuth Antimony Telluride Solid Solutions in Thermoelectrics Handbook. Macro to Nano, Ed. by D.M.Rowe, Taylor&Francis, 2006